%% file: cose-essay-svjour.tex
\documentclass[twocolumn]{svjour3}
\usepackage{svjour3fix}
\let\email=\relax
\usepackage[latin1]{inputenc}
\usepackage[colorlinks=true,ps2pdf=true]{hyperref}
\usepackage[dvips]{graphics}
\usepackage{psfrag}
\usepackage{amsmath}
\usepackage{amssymb}
\usepackage[english]{babel}
\usepackage[short]{bibstrings}
\usepackage{xspace}
\usepackage{url}
\usepackage{markup}

\date{\today}

\title{In Things We Trust? Towards trustability in the Internet of Things\thanks{%
  This research is supported by the research program
  Sentinels (\url{www.sentinels.nl}) as project 
  'Identity Management on Mobile Devices' (10522). 
  Sentinels is being financed by 
  Technology Foundation STW, the Netherlands Organization 
  for Scientific Research (NWO), and the Dutch Ministry of Economic Affairs.
}}

\author{Jaap-Henk Hoepman}
\institute{Jaap-Henk Hoepman \at
TNO (\email{jaap-henk.hoepman@tno.nl}), 
and\\
Institute for Computing and Information Sciences (ICIS),
Radboud University Nijmegen (\email{jhh@cs.ru.nl}).
}

\begin{document}

\maketitle

\bibliographystyle{acm}

\input cose-essay-body.tex

%
\bibliography{cose-essay}

\end{document}

%% file: cose-essay-body.tex
%

\section{Introduction}
\bodyversion{$Id: cose-essay-body.tex 10 2011-09-12 21:11:25Z jhh $}
The Internet of Things is nothing new. First introduced as Ubiquitous Computing by Mark Weiser~\cite{weiser1991ubicomp} around 1990, the basic concept of the ``disappearing computer'' has been studied as Ambient Intelligence or Pervasive Computing in the decades that followed. Today we witness the first large scale applications of these ideas.
We see RFID technology being used in logistics, shopping, public transport and the like. The use of smart phones is soaring. Many of them
are able to determine their location using GPS (Global Positioning System). Some phones already have NFC (Near Field Communication) capabilities, allowing them to communicate with objects tagged with RFID directly. Combined with social networking (like Facebook or Twitter), this gives rise to advanced location based services, and augmented reality applications. In fact social networks interconnecting things as well as humans have already emerged. Example are Patchube, a web-based service built to manage the world's real-time data\footnote{\url{https://pachube.com/}}
and Flukso, a web-based community metering application\footnote{\url{http://www.flukso.net/}}.

As the full ramifications of the Internet of Things start to unfold, this confluence of cyberspace and physical space is posing interesting new and fundamental research challenges. In particular, as we will argue in this essay, it has a huge impact in the area of security, privacy and trustability. As Bruce Schneier puts it in a recent issue of CryptoGram~\cite{schneier2011security-in-2020} (while discussing IT in general): 
\begin{quote}
``[...] it's not under your control, it's doing things without your knowledge and consent, and it's not necessarily acting in \emph{your} best interests.''
\end{quote}
The question then is how to ensure that, despite these adverse conditions, the Internet of Things is a safe, open, supportive and in general pleasant environment for people to engage with, or in fact for people to live in.

This essay is structured as follows. We define the Internet of Things in section~\ref{sec-vision}, and describe the main privacy, security and trustability issues associated with it in section~\ref{sec-problem}. Solutions to these problems will have to deal with certain constraints, as explained in section~\ref{sec-constraints}. Section~\ref{sec-thepast} discusses classical solutions based on data minimisation techniques, while section~\ref{sec-alt} discusses more recent alternative approaches. We conclude with an extensive overview of research challenges in section~\ref{sec-future}.

\section{The vision}
\label{sec-vision}

What exactly is the Internet of Things? Many definitions can be given. At a basic level the Internet of Things is a dynamic global network infrastructure with self configuring capabilities where physical and virtual ``things'' have identities, physical attributes, and virtual personalities. They use intelligent interfaces, and are seamlessly integrated into the information network~\cite{clusterbook2010}. Such ``things'' could be a pair of jeans (with an RFID tag attached), a light switch, a light bulb, a fridge, a washing machine, or any other sensor or actuator: the list of things is basically endless. All these things become first class members of the Internet, sharing their data with the world, and using the world's data for their own purposes.

Far more interesting is the envisioned applications of the Internet of Things to realise the Ambient Intelligence (AmI) concept. This concept
\begin{quote}
\ldots provides a vision of the Information Society future where the emphasis is on user friendliness, efficient and distributed services support, user-empowerment, and support for human interactions. People are surrounded by intelligent intuitive interfaces that are embedded in all kinds of objects and an environment that is capable of recognising and responding to the presence of different individuals in a seamless, unobtrusive and often invisible way.~\cite{ESTO2003AmI}
\end{quote}
In an ambient intelligence world created using the Internet of Things, devices work in concert to support people in carrying out their everyday life activities, tasks and rituals in easy, natural ways using information and intelligence that is hidden in the network connecting these devices~\cite{aarts2001ambient,istag2003AmI,greenfield2006everyware}.

Applications of Ambient Intelligence have been proposed in a wide variety of areas, like housing (home automation, smart washing machines), smart cities (sustainability and energy conservation), mobility (traffic management systems, congestion control, support for multi-modal transport, public transport ticketing), commerce (inventory management, marketing and advertising, store personalisation), education (digital libraries, digital museums), and health (self-treatment, long-distance monitoring)~\cite{ESTO2003AmI} to name but a few examples.

The Internet of Things may change the way we perceive the world completely. For one thing, the world around us will start to perceive us as well~\cite{gershenfeld2000}. The book you read may `read' you as well. How will that influence our relationship with the things around us? How will that influence our own self image?

\subsection{Properties of the IoT}
\label{ssec-props}

A pervasive system like the Internet of Things is characterised by the following system properties.
\begin{description}
\item[Invisible by design] A pervasive system pervades the human environment, but resides in the periphery or our attention. Pervasive devices are not explicitly there; they do not take up space on your desk, but are often integrated into other common objects like windows, doors or walls. They may not have a direct user interface, and may have limited computing, storage and power resources.
\item[Networked] Devices are interconnected by a seamless communication infrastructure, which is dynamic and massively distributed.
\item[Many-to-many] Devices do not have a 1-to-1 relationship with a user. Where a laptop and a mobile phone are personal devices used by one user, pervasive devices are not restricted to one person as a user. One person can use many pervasive devices, and one pervasive device can be used by many persons.
\item[Always on] Devices are always active, it is not necessary to first actively switch them on before any interaction can be had with the system.
\item[Distributed] The computing intelligence and effort of a pervasive system is not restricted to one device but is the combined computing effort of multiple devices. Pervasive systems are comprised of widely heterogeneous devices, and show emergent behaviour. 
\item[Context-aware] Pervasive systems have some knowledge of their context. They may, for example, be aware of other pervasive devices in their vicinity, or they may be able to measure location or temperature.
\item[Adaptive/spontaneous/autonomic] The information retrieved from sensors is used by a pervasive system to adapt its behaviour. This adaption is spontaneous, meaning that it is not triggered by a user pushing a button, but by more implicit actions of somebody, like for example entering a room.
\item[Natural human interface]
A pervasive system has an intuitive human computer interface. People should not need to think about how to interact with the system, as this should be natural, e.g. through speech, touch or movement.
\end{description}
With this understanding of the Internet of Things and its properties, we are ready to discuss the potential problems with the Internet of Things, and possible approaches to mitigate these problems.

\section{The problem}
\label{sec-problem}

The vision of the Internet of Things outlined above is certainly an attractive one. However, the very same components used to build this vision can also be used to create a totally different future. To prevent this vision to become our worst nightmare, basic guarantees have to be implemented that will protect our privacy and will maintain security. This will not happen without considerable effort, for the current trend in IT is detrimental to security and privacy. As Schneier puts it~\cite{schneier2011security-in-2020}: ``the boundary between inside and outside disappears (\emph{deperimeterization}), data is increasingly stored and treated in the cloud (\emph{decentralization}), general purpose computer is replaced by special purpose devices (\emph{deconcentration}), and smart software and devices will increasingly do things on our behalf (\emph{depersonization})''. We will describe the main privacy, security and trustability issues below.

\subsection{Privacy}

In a world of sensors and actuators that surround us and support us in our day to day activities, privacy is obviously a big concern.

Privacy --- sometimes loosely defined as the `right to be let alone'~\cite{warren1890privacy} --- is considered a fundamental human right in many societies. It is ``essential for freedom, democracy, psychological well-being, individuality and creativity''~\cite{solove2008understanding}.
Privacy has many dimensions (corporeal, relational, etc.), but for the purpose of this essay we focus on the data protection aspect of it. We wish to stress that data protection is not the same as keeping personal information confidential. Data protection laws and regulations are much broader.
They determine the conditions under which businesses and governments are allowed to collect, process and use personal information (proportionality and subsidiarity). They empower citizens to determine how personal data about them is used even after it is collected by third parties. They allow them to be informed about the use of their personal information, and give them the right to correct personal information about themselves. 

As a consequence, privacy protection in the Internet of Things~\cite{garfinkel2005rfid-privacy,juels2006rfid-secpriv-survey}
involves much more than data minimisation techniques like using pseudonyms and preventing data collection through proper access control. In fact, the vision of an Internet of Things that intelligently supports us in our day to day activities \emph{needs} to collect large amounts of personal information\ldots
The challenge is to accommodate this need for personal data, while maintaining privacy guarantees.

\subsection{Security}

Serious integrity, authenticity, and availability concerns arise too
in the Internet of Things.

Consider the use of RFID tags in supply chain management as an example. If the logistics of a company critically depends on the correct bookkeeping of items in stock through RFID tags, then inserting fake or wrong tags in the system can do serious damage. Radio interference or outright radio jamming may make inventory scanning impossible or highly inaccurate. Swapping tags on items in stock may allow customers to defraud store owners.  Recent research even indicates that (fake) RFID tags can be used to spread computer viruses~\cite{DBLP:conf/percom/RiebackCT06}.

When the Internet of Things expands to other application areas, like health care, smart grids, and the like, the Internet of Things itself becomes a critical infrastructure. This is especially the case when the nodes are not merely sensors but also actuators, whose actions control critical components.
This imposes strong security requirements. Not so much regarding confidentiality (although this is a concern with respect to industrial espionage related to supply chain information), but the more so regarding integrity, authenticity, and availability of the Internet of Things~\cite{juels2006rfid-secpriv-survey}. 

The issue also needs to be addressed at the management level. Who is in charge?
And when something goes wrong, who is responsible?~\cite{ESTO2003AmI} These questions are not so easily answered in a pervasive system like the Internet of Things where a single 'point of authority' is lacking.

\subsection{Trustability}

An even more principal issue, that partly underlies the security and privacy problems associated with the Internet of Things, is that of trust, or rather, trustability. In sociology, trust is defined as follows~\cite{hardin2002trust}
\begin{quote}
When an actor trusts another actor, she is willing to assume an open and vulnerable position. She expects the other to refrain from opportunistic behaviour even if there is the possibility to show this behaviour.
\end{quote}
Often designers of ICT infrastructure assume (or rather \emph{impose})
the need to trust the infrastructure by its users, because adequate privacy measures are missing, proper security is not guaranteed, and risks are not mitigated in any other way. A paradigm shift is needed away from this paternalistic 'trust us' implementation of the ICT infrastructure that surround us, to a more user-centric 'trustability' approach where the infrastructure allows the user to built up trust using personal tools and other means. We propose the following definition.
\begin{quote}
A system is \emph{trustable}, if the risk of using the system for a particular purpose can be reliably estimated by the user using third party tools under her own control, and/or using third party data of her own choosing.
\end{quote}
It is an interesting question how techniques from identity management (and solutions to its associated problems~\cite{alpar2011identitycrisis}), and the trusted computing paradigm~\cite{mitchell2005trusted-computing} can be re-applied in this new context.

\section{Constraints}
\label{sec-constraints}

The previous section has argued that strong privacy and security guarantees have to built in into the Internet of Things, in order to prevent disruptions in the scenarios outlined above. However, implementing these guarantees should not interfere with the realisation of the Internet of Things itself. This makes developing such guarantees an interesting research challenge, on which this essay will expound further. We note that the recommendation of the European Commission of May 2009~\cite{EC-2009-387} to kill RFID tags at the point-of-sale is a too disruptive in that respect: it strongly protects the privacy of the citizen, but makes it much harder to use RFID tags beyond the point-of-sale for all kinds of benevolent applications.

Classical security countermeasures and privacy enhancements do not apply to RFID due to their pervasiveness and limited computing power. Low cost
RFID tags do not have the resources to perform any but the most primitive cryptographic operations, and their sheer number pose scalability problems.
Similarly, new models, policies and assessment methodologies need to be developed: the linking of physical objects with the networked world
through RFID, and the new possibilities for profiling, pose new security and privacy threats that are not captured by the current state of the art. Solutions are further constrained by the properties of a pervasive system listed in section~\ref{ssec-props}.

\section{The past: data minimisation}
\label{sec-thepast}

Most research so far has focused on techniques to minimise data collection, by implementing certain forms of authentication and access control while respecting the resource constraints inherent to RFID based systems. We briefly review the state of the art in this area.

Early proposals use relabelling of tag identifiers~\cite{sarma2002rfid-report},
or re-encryption techniques~\cite{juels2003squealing,avoine2004rfidbanknotes,golle2004reencryption} that randomly encrypt the identifier from time to time, so that it can only be recovered by authorised readers, while being untraceable for others. 

Another approach is to implement some form of authentication between tag and
reader, and to allow only authorised tags to retrieve the tag identifier.
In a public key setting this would be easy, but RFID tags are generally
considered to be too resource poor to accommodate for that. 
Therefore, several identification and authentication protocols using hash functions or symmetric key cryptography have been proposed~\cite{weis2003security,engberg2004zeroknowledge-rfid}. 
In particular, Ohkubo, Suzuki, and Kinoshita~\cite{ohkubo2004hash-chains-rfid} present a technique for achieving forward privacy in tags. This property means that if an attacker compromises a tag, i.e., learns its current state and its key, she is nonetheless unable to identify the previous outputs of the same tag. In their protocol, a tag has a unique identifier $\id{id}_i$, that is changed every time the tag is queried by a reader. In fact, when queried for the $i$-th time, the tag responds with $g(\id{id}_i)$ to the reader, and sets 
$\id{id}_{i+1} = h(\id{id}_i)$ immediately after that. In both cases, if all readers are \emph{on line}, connected with one central database, the readers can be synchronised and the response of a tag can be looked up immediately in the database\footnote{%
  Note that the database can keep a shadow copy of $\id{id}_i$ and hence can
  precompute the next expected value $g(h(\id{id}_i))$.
}. 
If not, or if synchronisation errors occur, a search over all possible (initial) identifiers (expanding hash chains) is necessary.

In a symmetric key setting the reader cannot know the identifier of the tag
a priori, or obtain the identifier of the tag at the start of the protocol 
because of privacy concerns. One can give all readers and tags the same symmetric key, but this has the obvious drawback that once the key of one tag is stolen, the whole system is corrupted. To increase security,  tags can be given separate keys, but then the reader must search the right key to use for a particular tag. The core challenge is therefore to provide, possibly efficient, trade offs and solutions for key search and key management. Molnar and  Wagner~\cite{DBLP:conf/ccs/MolnarW04} (see also~\cite{DBLP:conf/percom/Dimitriou06}) propose to arrange keys in a tree
structure, where individual tags are associated with leaves in the tree, and
where each tag contains the keys on the path from its leaf to the root. In subsequent work Molnar, Soppera, and Wagner~\cite{DBLP:conf/sacrypt/MolnarSW05} explore ways in which the sub-trees in their scheme may be associated with individual tags. In another approach, 
Avoine, Dysli, and Oechslin~\cite{DBLP:conf/sacrypt/AvoineDO05,DBLP:conf/percom/AvoineO05} show how, similar to the the study of Hellman to breaking symmetric keys, a time-memory trade off can be exploited to make the search for the key to use
more efficient. We note that none of these systems are practical for RFID systems where millions of tags possess unique secret keys.

We refer to Juels~\cite{juels2006rfid-secpriv-survey} (and the excellent bibliography\footnote{\url{http://www.avoine.net/rfid/}} maintained by Gildas Avoine)
for a much more extensive survey of proposed solutions, and \cite{juels2007strong-privacy-rfid} for a more formal analysis of the
privacy properties actually achieved by some of the proposed authentication protocols.

\section{Alternative approaches}
\label{sec-alt}

Spiekermann~\etal\cite{spiekermann2009critical-rfid-pet}
observe that although there are many protocols and proposals for 
limiting access to RFID tags (either by killing them completely or by requiring
the reader to authenticate), few systems have been proposed that
allow effective and fine grained control over access
permissions. Recent research efforts have tried to bring the user back into control. Notable examples are agency tools like
the RFID Guardian~\cite{DBLP:conf/lisa/RiebackGCHT06} and the
Privacy Coach~\cite{hoepman2010privacycoach}, as well as
the ``resurrecting duckling''~\cite{StaA99} principle of Stajano and Anderson. 

\subsection{Design philosophies}

The ``resurrecting duckling''~\cite{StaA99} security policy model of Stajano and Anderson is an example of a general design philosophy applicable to the Internet of Things, that aims to put the user in better control of the devices that he owns or the devices that surround him. The principle is based in  analogy to the biological principle of \emph{imprinting} discovered by Lorentz~\cite{lorentz1949redete}, which describes the initial bonding process between hatched ducklings and their (supposed) parents. In this model a device is in two possible states: \emph{imprintable} or \emph{imprinted}. When imprintable, anyone can take ownership of the device. In doing so, the device becomes imprinted. Only the owner of an imprinted device may cause the device to 'die', bringing it back to its imprintable state (and resetting all other settings to default, essentially bringing the device back in a virgin, new-born, state). Additionally, an owner of a device may change security policies on the device, granting certain rights to other users. This allows an owner of a device to lend the device to another user, and delegate a subset of its power to this user.

More models like this need to be developed to better understand the nature of the Internet of Things. 

\subsection{Agency tools}

The RFID Guardian and the Privacy Coach can be classified as \emph{agency tools}: tools that support the user to make choices and to impose those choices on the world~\cite{bandura2001agentic}. Such tools put the user at the centre of the Internet of Things.

The RFID Guardian~\cite{DBLP:conf/lisa/RiebackGCHT06} is best understood as a personal firewall between the RFID tags carried by a user, and the world of RFID readers that surround the user. The user programs the RFID Guardian to grant or deny access to specific tags from certain readers, possibly depending on the current context. The RFID Guardian performs this task by selectively jamming radio signals if it detects a reader that tries to access a tag for which access is denied.

The Privacy Coach~\cite{hoepman2010privacycoach} puts the user in control in a different way. It is an application running on a mobile phone equipped with a reader that can read RFID tags. Certain such NFC enabled phones are currently on the market. The Privacy Coach supports users in making privacy decisions when confronted with RFID tags on items they buy (or otherwise obtain). It functions as a mediator between customer privacy \emph{preferences} and corporate privacy \emph{policies}, trying to find a match between the two, and informing the user of the outcome. 

The Privacy Coach itself does not block or prevent any privacy infringements. Instead, it stores the user privacy preferences in a profile on the mobile phone. Privacy policies associated with RFID tags are downloaded from a central database whenever the user scans such a tag using the NFC reader.
Producers of goods tagged by RFID will similarly store the company privacy policy associated with these tags in a central database. Alternatively, consumer organisations may create such privacy policies for companies that do not provide these policies themselves.

\section{Future challenges}
\label{sec-future}

The remainder of this essay is devoted to describing the main research challenges ahead.

\subsection{Privacy beyond data minimisation}

Current approaches to protect our privacy focus on data minimisation. This is as counterproductive in the Internet of Things as it is in social networks: both only 'work' if you are willing to share your data. This is not to say that in order for the IoT to be useful, your personal data needs to be shared with \emph{everybody}. Like in social networks, context separation~\cite{nissenbaum2004privacy} will play an important role in the Internet of Things as well. But simply refusing to share your data with anybody will not be possible (although in certain cases, anonymity mechanisms may still be applicable). 
 
This means that privacy enhancing technologies need to be developed that prevent the abuse of personal data once it is collected~\cite{hildebrandt2009transparency}, and that prevent the leakage of information from one context to the other (thus maintaining contextual integrity~\cite{nissenbaum2004privacy}). Design philosophies, and derived design patterns, for the Internet of Things need to be developed that accomplish this. Moreover, a common privacy engineering practise based on these principles needs to be established. These privacy preserving approaches need to be applicable to heterogeneous sets of devices~\cite{clusterbook2010}, and need to be user friendly. This adds to the research challenge already present. 

Several approaches can be followed to achieve this. One approach is to collect and maintain user profiles and preferences on a personal device held by the user (like a mobile phone) instead of by the infrastructure directly. 
The core data needed to make the ambient infrastructure intelligent is then still under control of the user. The infrastructure can query the user profile through standard interfaces provided by the personal device. In a way the personal device 
operates as a personal firewall. 
This approach is somewhat similar to recent studies into privacy enhanced profiling of website visitors. These techniques aim to implement targeted advertising on websites~\cite{DBLP:conf/fc/AndroulakiB10,toubiana2010adnostic} without the usually associated privacy problem of collecting user profiles centrally.

Alternatively, user profiles can be split into many small parts and stored at many different, uncorrelated, locations. This can even be done in such a way that wrong information is encoded in some of these parts. The parts can be combined using secret sharing techniques. In case of wilfully distributing wrong information, the wrong data can be filtered out using majority voting and other fault tolerant techniques~\cite{jalote1994fault-tolerance} once all the parts (correct and incorrect ones) have been collected.

\subsection{Security}

The main security properties that are relevant for the Internet of Things are
integrity, authenticity, and availability. These need to be achieved in an environment where the endpoints are mostly very resource constrained. Endpoints are typically tags, sensors and actuators, that need to be produced at the lowest possible cost (because a proper implementation of the Internet of Things will need so many of these nodes to be deployed). These endpoints have little memory, little processing power, and slow, short range and unreliable communication links. Security (and privacy) therefore need to be built upon resource efficient cryptographic primitives. This remains a challenging area of research.

Also, the Internet of Things will lack a single central authority. This calls for models for decentralised authentication~\cite{clusterbook2010}, including strategies for revocation and key-distribution in an ad-hoc fashion. 
In general, security measures need 
to support the conflicting requirements of multiple stakeholders (\eg privacy protection versus accountability), in order to support multilaterally secure cooperations~\cite{weber2009multilateral}, and should be designed in such a way that they can be used by casual users. This has to be achieved without the coordinating role of a central authority trusted by all stakeholders.

The same holds not only for devices. Reliability (or rather integrity) of the \emph{data} collected by the Internet of Things and provided back to the users is also an issue. Open source data mining tools to verify the reliability of the data may help in this respect. An example of an area where this is especially important is health care applications of the Internet of Things where patients share data to crowd-source knowledge about their diseases, and subsequently use that data to improve their standard over living\footnote{\cf \url{http://www.patientslikeme.com/}}.
The diffusion of harmful and unsubstantiated knowledge and information is a real possibility. However, experiences with similarly crowd-sourced knowledge bases like Wikipedia suggest that in an open system, malicious knowledge tends to gradually be muted out~\cite{storni2010iothealth}.

We note that security can also \emph{benefit} from the existence of an Internet of Things. Through the IoT it is much easier to reliably collect information about the context in which a certain actor tries to access a certain resource. The current location of the user, whether the user is alone in the room, whether someone else is approaching, whether certain devices are or are not in the vicinity: all these aspects can be determined. This allows us to specify much more fine grained access conditions, that can still be fulfilled given a much richer data set at the time the resource is accessed.

\subsection{Establishing trustability}

Establishing trust in the Internet of Things should go beyond the mere user perception side of the issue, but instead focus on measurable ways to establish trustability, and on tools to support this. Trustability
aims to answer questions like: How well does the infrastructure safeguard the data you entrust to it? What are the future consequences of its use? How clearly and openly do infrastructure providers advise you of your rights and responsibilities? What guarantees of future reliability and availability does the infrastructure give you? 

Very few of these tools exist to help the user to determine the trustability of the infrastructure it is engaging in. The issue is much more complex than simply determining whether a certain public terminal is authentic before entering your PIN code on it~\cite{DBLP:journals/cn/AsokanDSW99} (although certainly knowing the terminal is authentic helps to some extent). Methods based on direct anonymous attestation~\cite{DBLP:conf/ccs/BrickellCC04} using Trusted Platform Modules (TPM) (that establish that a certain device is a known good state) are of limited value. 
The sheer heterogeneity of the devices that make up the Internet of Things make it impossible to enumerate all the good states each of these devices can be in. Moreover, because the IoT has no central authority, and as context matters, the question is who to turn to to tell you what a good state of a certain device is in the first place. 

Most importantly though, establishing trust is a process, a process that progresses over time in which users adjust their trust assessment in the devices and infrastructures they engage in with every transaction they perform with them. The use of personal, mobile, devices and applications (\cf~\cite{weber2010trustworthy}) to support the user in this process need to be developed. These could for instance be used to predict the future consequences of current engagements with the Internet of Things 
(\cf\cite{hildebrandt2009transparency}). These ideas could build upon the results obtained in the Smart Products\footnote{\url{http://www.smartproducts-project.eu/}} project that aim to embed ``proactive knowledge'' into the IoT and consider e.g. usability and security in access control mechanisms based on machine learning techniques to make the configuration of the IoT manageable by casual users.

Transparency is a key factor in the aforementioned process. Transparency helps the user to assess the trustability of a party in a transaction. It also provides useful meta-information that is essential to establish the integrity of the data collected by the IoT. We therefore need to engineer built-in transparency for the IoT, and develop the concept of \emph{transparency by design} similar to privacy by design.

\subsection{Governance}

\begin{quote}
[...] the more autonomous and intelligent 'things' get, problems like the identity and privacy of things, and responsibility of things in their acting will have to be considered~\cite{clusterbook2010}.
\end{quote}

Governance can be defined as ``the use of institutions and structure of authority to allocate resources and coordinate or control activity in society''~\cite{bell2002governance}. The three main stakeholders (government, the private sector and the civil society) should be represented in these institutions and structures of authority. But what these institutions should be and what this structure of authority should look like is currently unclear for the Internet of Things\footnote{Private communication, from the Internet of Things Expert Group~\cite{EC-IoT-expert-group}}. 

It is pretty much a chicken-and-egg problem. 

Because there is no common view on the future and design of the Internet of Things, it is hard to define an appropriate governance for it. As particular case in point, it has been observed that things are bound to physical locations. It is therefore foreseen that the Internet of Things will have a much more localised nature than the current Internet. In fact there may not even be a single Internet of Things. Instead, there may several networks of things, perhaps each using different technology, operating as pretty much isolated islands of interconnected things. 

On the other hand, because of a lack of governance, there is no (visible and accountable) converging force that will slowly bring together the different views and designs for the Internet of Things. This lack of transparency and openness may have a negative impact on the acceptance of this new technology in our society. Especially because the consequences of this new technology are quite radical. 

This chicken-and-egg problem has to be resolved, because governance cannot be retrofitted. The history of the development of the Internet itself may serve as an example. Even though the Domain Name System (DNS) works, for better or worse, from a technical perspective, it has severe legitimacy problems because of decisions made early on that did not foresee the development of the Internet as it is now. Trying to change the governance structure today proves to be very difficult because of vested commercial and governmental interests. 

When setting up a governance structure care has to be taken not to overdo it. The very power of the Internet, that made it grow as fast as it did, is the almost 'anarchistic' nature of the underlying technology~\cite{lessig1999code}. This has ensured that no single party can control the whole network, and that all types of traffic are treated equal.

\section{Acknowledgements}

I would like to thank the members of Council\footnote{\url{http://www.theinternetofthings.eu}} (Rob van Kranenburg, Erin Anzelmo, Cristiano Storni, James Wallbank, Tijmen Wisman) and the members of IFIP WG 11.2\footnote{\url{http://www.cs.ru.nl/ifip-wg11.2}} (Denis Trcek, Stefan Georg Weber, Igor Ruiz-Agundez) for their input to this essay.

\section{Biography}

Jaap-Henk Hoepman is senior scientist at the security group of TNO, the Dutch Organisation for Applied Scientific Research. He also holds an associate professor position at the Digital Security group of the Institute for Computing and Information Sciences of the Radboud University Nijmegen.
He is chair of IFIP WG 11.2 on Pervasive Systems Security, and member of Council (a multidisciplinary think tank on the Internet of Things).